\documentclass[12pt,preprint]{aastex}

\shorttitle{Cool and Warm IR Galaxies}
\shortauthors{Rahman, Helou, \& Mazzarella}
\begin{document}

%%%%%%%%%%%%%%%%%%%%%%%%%%%%%%%%%%%%%%%%%%%%%%%%%%%%%%%%%%%%%%%%%%%%%%%%%%%%%%%%
\title{Probing Cool and Warm Infrared  Galaxies using Photometric and 
Structural Measures}
%%%%%%%%%%%%%%%%%%%%%%%%%%%%%%%%%%%%%%%%%%%%%%%%%%%%%%%%%%%%%%%%%%%%%%%%%%%%%%%%

\author{Nurur Rahman\altaffilmark{1}, George Helou, and Joseph M. Mazzarella}
\affil{Infrared Processing and Analysis Center/Caltech, \\
Mail Code 100-22, 770 S. Wilson Avenue, \\ 
Pasadena, CA 91125, USA}
\email{nurur@ipac.caltech.edu; gxh@ipac.caltech.edu; mazz@ipac.caltech.edu} 

\altaffiltext{1}{National Research Council (NRC) Postdoc Fellow}
%%%%%%%%%%%%%%%%%%%%%%%%%%%%%%%%%%%%%%%%%%%%%%%%%%%%%%%%%%%%%%%%%%%%%%%%%%%%%%%%
%%%%%%%%%%%%%%%%%%%%%%%%%%%%%%%%%%%%%%%%%%%%%%%%%%%%%%%%%%%%%%%%%%%%%%%%%%%%%%%%
%%%%%%%%%%%%%%%%%%%%%%%%%%%%%%%%%%%%%%%%%%%%%%%%%%%%%%%%%%%%%%%%%%%%%%%%%%%%%%%%
%%%%%%%%%%%%%%%%%%%%%%%%%%%%%%%%%%%%%%%%%%%%%%%%%%%%%%%%%%%%%%%%%%%%%%%%%%%%%%%%
\begin{abstract}
We have analyzed a sample of nearby cool and warm infrared (IR) galaxies 
using photometric and structural parameters. The set of measures include 
far-infrared color ($C = \log_{10}[S_{60\mu m}/S_{100\mu m}]$), total IR 
luminosity ($L_{TIR}$), radio 
surface brightness as well as radio, near-infrared, and optical sizes.
In a given luminosity range cool and warm galaxies are considered as those 
sources that are found approximately $1 \sigma$ below and above 
the mean color in the far-infrared $C - L_{TIR}$ diagram. 
We find that galaxy radio surface brightness is well correlated with color 
whereas size is less well correlated with color. 
Our analysis indicates that IR galaxies that are dominated by cool dust are 
large, massive spirals that are not strongly interacting or merging and 
presumably the ones with the least active star formation. Dust in these cool 
objects is less centrally concentrated than in the more typical luminous and 
ultra-luminous IR galaxies that are dominated by warm dust. 
Our study also shows that low luminosity early type unbarred and transitional 
spirals are responsible for the large scatter in the $C - L_{TIR}$ diagram. 
Among highly luminous galaxies, late type unbarred spirals are predominately 
warm, and early type unbarred and barred are systematically cooler. 
We highlight the significance of $C - L_{TIR}$ diagram in terms of local and 
high redshifts sub-millimeter galaxies.
\end{abstract}
%%%%%%%%%%%%%%%%%%%%%%%%%%%%%%%%%%%%%%%%%%%%%%%%%%%%%%%%%%%%%%%%%%%%%%%%%%%%%%%
%%%%%%%%%%%%%%%%%%%%%%%%%%%%%%%%%%%%%%%%%%%%%%%%%%%%%%%%%%%%%%%%%%%%%%%%%%%%%%%
%%%%%%%%%%%%%%%%%%%%%%%%%%%%%%%%%%%%%%%%%%%%%%%%%%%%%%%%%%%%%%%%%%%%%%%%%%%%%%%
%%%%%%%%%%%%%%%%%%%%%%%%%%%%%%%%%%%%%%%%%%%%%%%%%%%%%%%%%%%%%%%%%%%%%%%%%%%%%%%
%%%%%%%%%%%%%%%%%%%%%%%%%%%%%%%%%%%%%%%%%%%%%%%%%%%%%%%%%%%%%%%%%%%%%%%%%%%%%%%
%%%%%%%%%%%%%%%%%%%%%%%%%%%%%%%%%%%%%%%%%%%%%%%%%%%%%%%%%%%%%%%%%%%%%%%%%%%%%%%
%%%%%%%%%%%%%%%%%%%%%%%%%%%%%%%%%%%%%%%%%%%%%%%%%%%%%%%%%%%%%%%%%%%%%%%%%%%%%%%
%%%%%%%%%%%%%%%%%%%%%%%%%%%%%%%%%%%%%%%%%%%%%%%%%%%%%%%%%%%%%%%%%%%%%%%%%%%%%%%
\section{Introduction}
Observations by the {\it Infrared Astronomical Satellite} ({\it IRAS}) led to the 
identification  of very luminous and ultra-luminous infrared galaxies, 
commonly known as the LIRGs and ULIRGs, respectively. These galaxies have 
enormous far-infrared (FIR) luminosities, 
$10^{11} < L_{FIR}/L_{\odot} < 10^{12}$ for LIRGs and 
$L_{FIR} > 10^{12} \ L_{\odot}$ for ULIRGs, and emit the bulk of their energy 
at the infrared wavelengths (Soifer et al. 1984, 1986, 1987; Sanders et al. 
1988). Subsequent multi-wavelengths studies reveal that in the local universe 
($z < 0.1$) ULIRGs have comparable total luminosities ($L_{TIR}$) and 
higher space densities than those of optical quasars (Sanders et al. 1988; Kim 
\& Sanders 1998). 

Understanding the nature and origin of energy sources in the LIRGs and ULIRGs 
has been the subject of much debate. Studies indicate that in the vast majority 
of these objects power comes from dust which is heated to various temperatures 
by various thermal and non-thermal processes such as ongoing (steady-state) 
star formation, intense starburst phase, and/or synchrotron radiation from the 
supernovae explosions (Lawrence et al. 1986; Genzel et al. 1998; Lutz et al. 
1998). These systems may also be powered by active galactic nuclei (AGN; 
Lonsdale, Smith, \& Lonsdale 1993). 
The energy distributions suggest that most of the observed FIR emission from  
luminous disk galaxies is due to the thermal radiation from warm dust heated by 
hot stars embedded in HII regions and molecular clouds heated directly by young 
OB stars; and cool dust from the general interstellar radiation field (Helou 
1986; Presson \& Helou 1987; Barvainis, Antonucci, \& Coleman 1992). Dust could 
also be heated by shocks in the interstellar medium during the collision or 
interaction of galaxies (Harwit et al. 1987). 

Dust heating mechanisms can be traced indirectly as each of these should 
correspond to particular spatial distributions of the FIR light. 
For example, the FIR emission originated by galaxies dominated by an AGN would 
appear compact and unresolved.
On the other hand, starburst heated dust should have about the same scale size 
as the burst itself because young stars are well mixed with the gas from which 
stars are forming. 
Dust heated  by the non-ionizing photons from cold, population 
I stars, might be expected to follow the smoothed distribution of older giants 
in the galaxy (Zink et al. 2000). A determination of the total energy output 
such as luminosity, the amount of dust present in the interstellar medium and 
its spatial distribution within the galaxies as well as its relationship to 
other basic components, such as the atomic and molecular gas, the stars, etc, 
is essential to studies connecting galactic structure and nature of the IR 
energy sources (Carico et al. 1990; Sopp \& Alexander 1991, 1992; Andreani \& 
Franceschini 1992, 1996; Alton et al. 1998; Domingue et al. 1999; Siebenmorgen, 
Kr\"{u}gel \& Chini 1999; Haas et al. 2000; Stickel et al. 2000; Trewhella et 
al. 2000; Zink et al. 2000).

The FIR color $C$, defined by the {\it IRAS} $60$ -to- $100$ $\mu m$ flux 
density ratio, an indicator of characteristic dust temperature, is a 
diagnostic of the typical heating conditions in the interstellar medium of a 
galaxy (Bothun, Lonsdale, \& Rice 1989; Soifer \& Neugebauer 1991).  It has 
been shown that low-redshift {\it IRAS} galaxies exhibit positive correlation 
of $C$ with $L_{FIR}$ such that more luminous galaxies tend to be warmer as 
compared to less luminous galaxies (Soifer et al. 1987; Dale et al. 2001; 
see Fig.1 in this study). The color changes systematically over roughly 
three orders of magnitude in luminosity with large scatter. As a result 
there is a substantial number of highly luminous but cool galaxies as well 
as low luminosity yet warm/hot galaxies (Dale et al. 2001; Chapman et al. 
2003). 
In this study we attempt to understand cool and warm IR luminous sources in 
terms of the geometric distribution of dust and galaxy morphology. 
In a given FIR luminosity range, these galaxies are considered in the context 
where cool galaxies are $1 \sigma$ below and warm galaxies are $1 \sigma$ 
above the mean color at a given luminosity. 
We are interested in differences between cool and warm galaxies in the same 
luminosity range that might suggest differences in the nature of the heating 
sources.

In an interesting study Young (1999) showed that star formation efficiency 
(SFE), defined by the ratio of $L_{TIR}$ to molecular hydrogen mass $M_{H_2}$, 
anti-correlated with disk (optical) sizes. The trend is prevalent in 
galaxies of various Hubble types and environments. It has been attributed to 
the shear present in disk galaxies. The molecular clouds in larger disks would 
experience increased turbulence which would reduce the efficiency of star 
formation (Young 1999). Galaxies in Young's sample span a broad range in FIR 
color comprising normal star forming galaxies (SFGs, 
$L_{FIR}/L_{\odot} < 10^{11}$) as well as LIRGs and ULIRGs. In Young's study, 
however, it remains unexplored how cool and warm systems behave in the visual 
as well as in the longer wavelengths. Since the IR luminous galaxies contain 
very large amount of dust, the interpretation of any observation at optical or 
shorter infrared wavelengths gets complicated by extinction. Therefore, to 
probe the origin of large IR luminosity it is necessary to have the knowledge 
of the spatial extents of the emitting regions in observations unaffected by 
dust extinction.

The paper is organized as follows: we describe the data in $\S2$ and parameter 
error estimation in $\S$3. We present our results in $\S$4 and discuss the 
implications of our results in $\S$5. The conclusions are given in $\S$6.  
%%%%%%%%%%%%%%%%%%%%%%%%%%%%%%%%%%%%%%%%%%%%%%%%%%%%%%%%%%%%%%%%%%%%%%%%%%%%%%%%%%%%%%
%%%%%%%%%%%%%%%%%%%%%%%%%%%%%%%%%%%%%%%%%%%%%%%%%%%%%%%%%%%%%%%%%%%%%%%%%%%%%%%%%%%%%%
%%%%%%%%%%%%%%%%%%%%%%%%%%%%%%%%%%%%%%%%%%%%%%%%%%%%%%%%%%%%%%%%%%%%%%%%%%%%%%%%%%%%%%
%%%%%%%%%%%%%%%%%%%%%%%%%%%%%%%%%%%%%%%%%%%%%%%%%%%%%%%%%%%%%%%%%%%%%%%%%%%%%%%%%%%%%%
%%%%%%%%%%%%%%%%%%%%%%%%%%%%%%%%%%%%%%%%%%%%%%%%%%%%%%%%%%%%%%%%%%%%%%%%%%%%%%%%%%%%%%
%%%%%%%%%%%%%%%%%%%%%%%%%%%%%%%%%%%%%%%%%%%%%%%%%%%%%%%%%%%%%%%%%%%%%%%%%%%%%%%%%%%%%%
%%%%%%%%%%%%%%%%%%%%%%%%%%%%%%%%%%%%%%%%%%%%%%%%%%%%%%%%%%%%%%%%%%%%%%%%%%%%%%%%%%%%%%
%%%%%%%%%%%%%%%%%%%%%%%%%%%%%%%%%%%%%%%%%%%%%%%%%%%%%%%%%%%%%%%%%%%%%%%%%%%%%%%%%%%%%%
\section{Data}
$IRAS$ resolved only the largest and nearest galaxies because of its 
comparatively large beam size of $\sim 2^{'}-4^{'}$ at 60 and 100 $\mu m$. 
Therefore, the FIR brightness distributions of most of the sources detected 
by the $IRAS$ are unknown. In contrast to the rather large scatter in 
optical-FIR correlations, a tight correlation of FIR with radio continuum 
total flux densities for infrared-selected galaxies appear to hold locally 
within individual galaxies (Helou, Soifer \& Rowan-Robinson 1985; Beck \& 
Golla 1988; Bicay, Helou, \& Condon 1988; Murphy et al. 2006). 
In nearby galaxies where ($IRAS$) 
FIR sizes have been measured directly, radio sizes are known to match, or 
to be somewhat larger than the size scale in FIR (Bicay \& Helou 1990). 
Besides, FIR and radio continuum brightness distributions of these galaxies 
show remarkable similarity (Marsh \& Helou 1995). 
The agreement in FIR and radio sizes, albeit derived from a smaller sample, 
suggest that high resolution radio maps may be taken as a good substitute, 
or at least as an upper limit for the size of the unobtainable FIR maps. 
We, therefore, use 1.49 GHz radio size as a proxy for the size of FIR emission.  

Wang \& Helou (1992, hereafter WH) studied the compactness of FIR bright 
galaxies and constructed a list of 330 galaxies from the $IRAS$ Bright Galaxy 
Sample (BGS; Soifer et al. 1986, 1987, 1989). WH constructed a flux limited 
sample of extragalactic objects brighter than 5.24 Jy at 60 $\mu m$, covering 
the entire sky surveyed by the $IRAS$ at galactic latitude $|b| > 30^{o}$. 
For radio data (rest frame flux density, size) WH used the atlas of the 1.49 
GHz radio maps based on VLA observations of IR bright galaxies that had been 
compiled by Condon, Anderson \& Helou  (1990, hereafter CAH). After removing 
AGN dominated galaxies identified as highly compact radio sources, the final 
list of WH contained 218 galaxies. Our sample is based on this list of 218 
galaxies. We use the same radio data, however, we take the rest frame $IRAS$ 
flux densities, luminosity, and estimated distance of member galaxies from the 
Revised Bright Galaxy Sample (RBGS) since it provides the best available 
reference for accurate $IRAS$ fluxes and IR luminosities of galaxies in the 
local universe (see Sanders et al. 2003 for details on RGBS).

Particular attention is given to exclude AGN from the sample. Although the list 
of WH was carefully compiled to get rid of the radio monsters, we have checked 
the entire sample following the prescription of de Grijp et al. (1985) using mid 
and far infrared color - color correlations. We have found 5 galaxies that fall 
into the region of the color - color plot which is mostly occupied by AGN like  
sources (de Grijp et al. 1985). 
We have removed them from the list. The radio fluxes of the remaining 
213 galaxies correlate well with the IRAS fluxes, and all galaxies fall within 
$\sim 1.5 \sigma$ of the mean $q$-value in the $L_{FIR} - q$ correlation 
indicating absence of radio excess objects ($q$ parameter is defined in Condon, 
Anderson \& Helou 1991). We have also made visual inspection of the observed 
spectral energy distributions of member galaxies. From this analysis we conclude 
that our final sample of 213 sources are star forming disk galaxies, although 
there may be AGNs present that are not energetically dominant in the far-infrared.

The number of SFGs, LIRGs, and ULIRGs in our sample are 154, 52, and 7, 
respectively. These three classes of galaxies have the following redshift 
distributions: $ 0.0< z <0.016$, $0.012< z <0.051$, and $0.018< z <0.082$ 
with the median redshifts $0.006, 0.023,$ and 0.055. 
The effective color temperature of the integrated dust emission in the sample 
galaxies ranges in between $\sim 25K - 40K$, assuming emissivity index, 
$\beta = 2$ (Dunne et al. 2000; Dunne \& Eales 2001). 
The ULIRGs show signs of interaction, e.g., presence of 
neighbor and accretion, e.g., tidal tails and disturbed outer envelopes. 
Majority of LIRGs appear as single, isolated systems while $30\%$ of them 
appear disturbed. At the lower end of the luminosity range, the LIRGs 
are luminous single isolated galaxies, with features resembling to the low 
luminosity SFGs. A minority of SFGs ($15\%$) has disturbed outer envelope 
or nearby companion. 

The NASA$/$IPAC Extragalactic Database (NED) and the Lyon-Meudon Extragalactic 
Database (LEDA) has been used to obtain Galactic extinction corrected $B$-band 
luminosity $L_B$ and 25 mag arcsec$^{-2}$ linear diameter $D_B$. 
The near-infrared (NIR) $K_s$ band ($2.17 \mu m$) magnitude of our sample is 
$5.0 < K_s < 10.6$.
For bright galaxies Jarrett et al. (2000) have recommended $K_s$ band 20 mag 
arcsec$^{-2}$ diameter $D_{NIR}$ as the most reliable and robust for galaxy 
photometry. We obtain $D_{NIR}$ from Two Micron All Sky Survey catalog 
(Jarrett 2000) using a $5^{''}$ search radius. We did not find any reliable 
$B$-band size for CGCG247-020, IRASF 08339+6517, and IRASF 12132+5313, and 
NIR size for NGC 5256, NGC 5331, and MCG+07-23-019. All these galaxies belong
to the LIRGs sub-sample.     
%%%%
No correction was applied to the luminosities and diameters for internal 
extinction or inclination of the host galaxies. Following RBGS we adopt 
$\Omega_M = 0.3$, 
$\Omega_{\Lambda} = 0.7$, and $H_0 = 75$ km sec$^{-1}$ Mpc$^{-1}$. These 
values are slightly different than recent estimates from Wilkinson Microwave 
Anisotropy Probe (WMAP; Spergel et al. 2003). Table 1 includes a partial list 
of low luminosity star forming galaxies showing only cool and warm sources. 
In the table galaxies are sorted in ascending order of FIR color, i.e. galaxy 
numbered 1 has the lowest FIR color, for all luminosity classes
[{\it See the electronic edition of the Journal for the complete list of galaxies}]. 
%%%%%%%%%%%%%%%%%%%%%%%%%%%%%%%%%%%%%%%%%%%%%%%%%%%%%%%%%%%%%%%%%%%%%%%%%%%%%%%%%%%%%
%%%%%%%%%%%%%%%%%%%%%%%%%%%%%%%%%%%%%%%%%%%%%%%%%%%%%%%%%%%%%%%%%%%%%%%%%%%%%%%%%%%%%
%%%%%%%%%%%%%%%%%%%%%%%%%%%%%%%%%%%%%%%%%%%%%%%%%%%%%%%%%%%%%%%%%%%%%%%%%%%%%%%%%%%%%
%%%%%%%%%%%%%%%%%%%%%%%%%%%%%%%%%%%%%%%%%%%%%%%%%%%%%%%%%%%%%%%%%%%%%%%%%%%%%%%%%%%%%
\section{Uncertainty in Parameters}
We use $L_{TIR}$ as a photometric measure, instead of $L_{FIR}$, since the 
former is based on the all four fluxes measured by the $IRAS$ and thus taking 
contributions from almost the entire IR range (Sanders \& Mirabel 1996). 
The median uncertainties associated with the $IRAS$ flux densities are 
$\sim 5\%$ and $\sim 3\%$ at 12 and 25 microns, respectively, and $\sim 1\%$ 
in both far-infrared bands. This leads to a $2\%$ uncertainty in the FIR color 
of a galaxy which is much smaller than the spread in mean color in a given 
luminosity.
The total IR flux of a galaxy is a weighted sum of four $IRAS$ flux densities. 
With the corresponding weighting factors given in Sanders \& Mirabel (1996) 
the median uncertainty in the total IR flux of a galaxy is $\sim 1\%$. 

Distances for the galaxies are taken from the RBGS. Most of the RBGS distance 
estimates come from redshift measurements, application of the Hubble law and 
correction for the Mould et al. (2000) flow model. However, some are 
primary (P) or secondary (S) distance estimates as flagged in the RBGS Table 1 
which do not come from the Hubble flow and cosmic attractor model. 
Our sample contains 48 galaxies of such distance estimates, all of which belong 
to the SFGs sub-sample. 
%Distances estimates for the galaxies are taken from the RBGS. They are derived 
%from the observed heliocentric radial velocity after correcting for non-Hubble 
%flow due to local over-densities following the the cosmic attractor model of 
%Mould et al. (2000). 
A significant uncertainty to a galaxy size and $L_{TIR}$ would result from an 
uncertainty in the distance estimate. Surface brightness being the distance 
independent measure would not be affected. The observed heliocentric radial 
velocities of sample galaxies are measured $\sim 1-2\%$ accuracy (obtained 
from NED). The uncertainties associated with the model parameters such as 
motions in the Local group, Virgo centric infall, Great Attractor infall etc. 
are high ($\geq 10\%$; Mould et al. 2000). The uncertainty associated with 
other distance estimates is of similar magnitude.
A galaxy optical angular size has less than $10\%$ uncertainty, depending on 
the flattening of the disk, as given in the Third Reference Catalog 
(de Vaucouleures et al. 1991; RC3). We assume a similar uncertainty in the 
NIR and radio bands (we note that it varies in different bands). Combining 
all of the preceding, we assign $\sim 15\%$ and $\sim 20\%$ uncertainty, 
respectively, to the estimate of physical size and $L_{TIR}$ of a galaxy. 

The uncertainty in $1.49 \ GHz$ surface brightness is due to the measurement 
errors in 1.49 GHz flux density and angular size. According to CAH (1990), 
the contributions from rms confusion error and calibration error to the 
$1.49 \ GHz$ flux density are smaller than the noise error.
The rms radio map noise $\sigma_n$ is between 0.1 and 0.2 mJy per beam solid 
angle $\Omega_b \sim \theta^2_{beam}$ (Condon 1987; CAH 1990). 
The noise contribution to the flux density is 
$\sigma_S \approx \sigma_n (\Omega/\Omega_b)^{1/2}$ in mJy, where 
$\Omega \sim \theta^R_M \times \theta^R_m$ is the solid angle covered by the 
source and $\theta^R_M $, $\theta^R_m$ are the deconvolved major ($M$) and 
minor ($m$) axes of a radio image.
Radio maps in CAH (1990) ranges in FWHM angular resolution from $1.5^{''}$ 
to 60$^{``}$ depending on the source's apparent size. Using $\sigma_n = 0.2$ 
mJy for each galaxy gives an uncertainty of $\sim 2\%$ in $1.49 \ GHz$ flux 
density. 
Uncertainty in radio size varies with beam resolutions but it is within $5\%$ 
in all cases. Taking this upper limit we find $\sim 10\%$ uncertainty in the 
surface brightness coming from $1.49 \ GHz$ flux density and angular size.   
%%%%%%%%%%%%%%%%%%%%%%%%%%%%%%%%%%%%%%%%%%%%%%%%%%%%%%%%%%%%%%%%%%%%%%%%%%%%%%%%%%%%%
%%%%%%%%%%%%%%%%%%%%%%%%%%%%%%%%%%%%%%%%%%%%%%%%%%%%%%%%%%%%%%%%%%%%%%%%%%%%%%%%%%%%%
%%%%%%%%%%%%%%%%%%%%%%%%%%%%%%%%%%%%%%%%%%%%%%%%%%%%%%%%%%%%%%%%%%%%%%%%%%%%%%%%%%%%%
%%%%%%%%%%%%%%%%%%%%%%%%%%%%%%%%%%%%%%%%%%%%%%%%%%%%%%%%%%%%%%%%%%%%%%%%%%%%%%%%%%%%%
\section{Results}
\begin{figure}
\epsscale{0.85}
\plotone{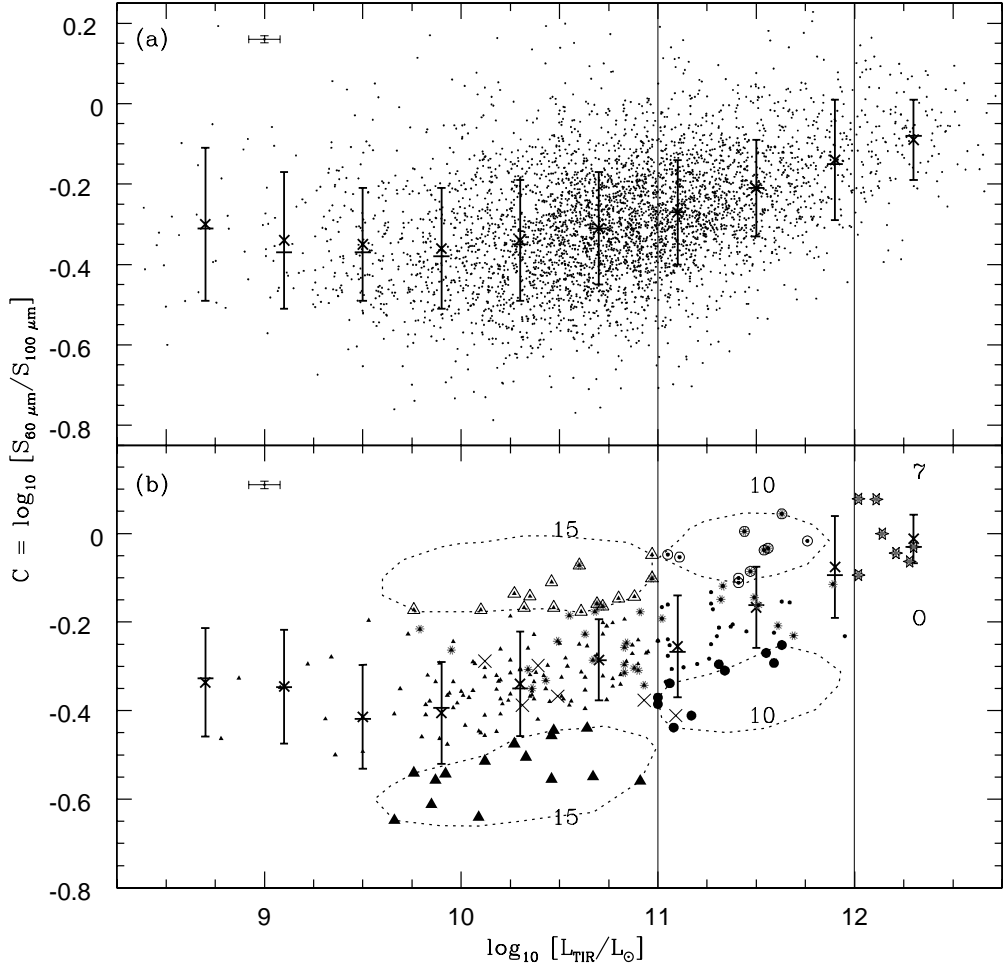}
\caption{FIR color as a function of $L_{TIR}$ for the $IRAS$ $S_{60 \mu m} > 1.2$ 
Jy sample of $\sim 4350$ galaxies (panel a), and our sample of 213 galaxies 
(panel b). Three classes of galaxies, in general, are shown by small triangles 
(SFGs), small dots (LIRGs), and stars (ULIRGs). 
The cool and warm sources are indicated, respectively, by large filled symbols 
and open symbols (triangles for SFGs and circles for LIRGs) in the regions 
delineated by dotted closed curves. In each luminosity bin the horizontal 
bar and (small) cross represent median and mean color. 
Each luminosity class (delineated by long vertical lines) contains equal number 
of cool and warm sources: 15+15 for SFGs and 10+10 for LIRGs. 
Large crosses and asterisks represent, respectively, local sub-mm and 
interacting/disturbed galaxies. A representative error bar is shown at the top 
of each panel.   
\label{colufig}}
\end{figure}

Figure \ref{colufig} shows the FIR color - IR luminosity diagram for $IRAS$ 
$S_{60 \mu m} > 1.2$ Jy sample of $\sim$ 4700 sources (Fisher et al. 1995) 
in panel a and our sample of 213 galaxies with $S_{60 \mu m} > 5.4$ Jy in 
panel b. The former sample is shown after removing 330 spurious cold luminous 
galaxies as pointed out by Chapman et al. (2003). In both panels horizontal 
bar and small cross represent median, and mean color with $1 \sigma$ error 
bar. These panels highlight the fact that galaxy FIR color and total IR 
luminosity follow a broad correlation with large spread in different flux 
limited samples.   
In this figure small triangles and dots represent SFGs and LIRGs, 
respectively (Fig.\ref{colufig}b). For these two classes of galaxies solid 
symbols represent cool sources (around or below $1 \sigma$ from the mean $C$) 
and open symbols represent the warm sources (around or above $1 \sigma$ from 
the mean $C$. Open stars are used for the ULIRGs. Because of their limited 
number we are unable to separate these ultra-luminous galaxies into cool 
and warm categories. In this panel large crosses represent local sub-milimeter 
galaxies (see below for a discussion of these galaxies).

\begin{figure}
\epsscale{0.85}
\plotone{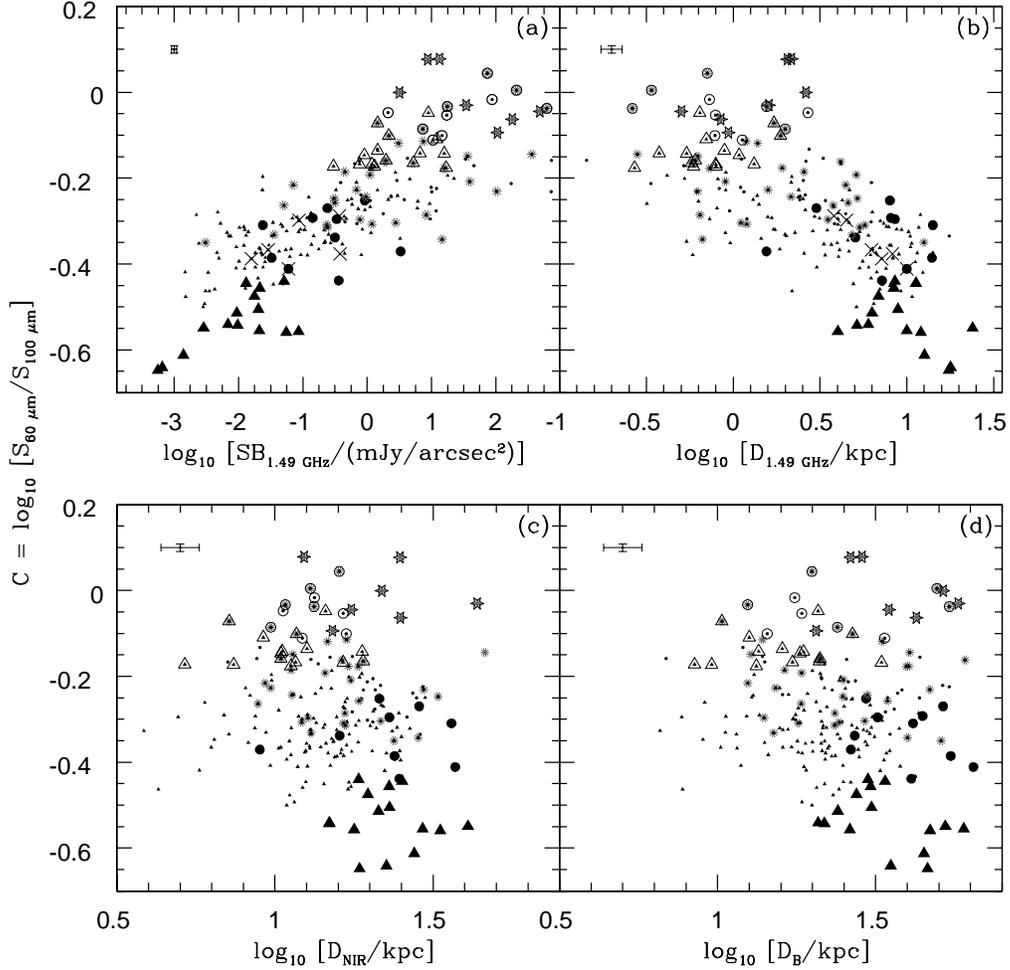}
\caption{FIR color as a function of radio surface brightness (panel a), radio 
size (panel b), NIR size (panel c), and optical size (panel d). 
In each luminosity class cool sources have smaller surface brightness than the 
warmer ones (panel a). The FIR/Radio emission in the cool systems are extended 
over larger area compared to the warmer one where the emission is more compact 
(panel b). A weak trend between cool and warm sources can also be noticed in 
the NIR and visual bands (panels c and d). Large crosses (in panels a and b) 
and asterisks represent, respectively, local sub-mm and interacting/disturbed 
galaxies. The symbols used are the same as in Fig. \ref{colufig}. 
A representative error bar is shown at the top left in each panel. 
\label{mainfig}}
\end{figure}

Figure \ref{mainfig} shows color as a function of SB$_{1.49 GHz}$ as well as 
galaxy size in radio, NIR, and optical bands. The choice of these parameters 
are based on conventional practice which suggests that radio size is a 
stand-in for FIR emission size, NIR size represents emission from old stellar 
population, and optical size accounts for the emission from a mixture of both 
young and old stars but complicated by dust. We have used the major axis to 
estimate the area of each object 
%($A=\pi a^2$ instead of $A=\pi a b$) 
to reduce the effects of highly uncertain inclinations. The radio surface 
brightness is calculated as $S_{1.49 GHz}/(\theta_M/2)^2$.
% flux density to the square of radio major axis. 
%A galaxy's physical size is 
%measured using isophotal major axis and SB$_{1.49 GHz}$ is calculated as 
%the ratio of $1.49 \ GHz$ flux density to the square of radio major axis. 
%None of these physical parameters has been corrected for inclination. 
%The inclination angle of sample galaxies ranges in between $\sim 20^{o}$ to 
%almost edge on. The use of major axis in size estimates reduces the effect 
%due to inclination. 

Cool and warm galaxies separate clearly in the surface brightness vs. color 
diagram (Fig.\ref{mainfig}a). For each luminosity class, warm colors 
correlate with high surface brightness. 
Panel b indicates that galaxies that are dominated by cool dust are large, 
massive spirals that are not strongly interacting or merging and presumably 
the ones with the least active star formation.
Dust in these cool objects is also, on average, less centrally concentrated 
than in the more typical LIRGs and ULIRGs that are dominated by warm dust. 
There is a clear sequence in the progression 
of color with radio size: color increases systematically from normal/isolated 
disks toward merger/disturbed disks. The latter type of disks are shown by 
asterisk symbols. Panels a and b shows that color is strongly correlated with 
both surface brightness and galaxy size at $1.49 \ GHz$ (see Table 2 for 
correlation statistics). 
Note that the uncertainty in color or brightness is statistical in nature 
and do not include calibration error. However, it will not affect our results 
since incorporating calibration error will simply shift these physical 
parameters systematically.

The statistical trend in Fig.\ref{mainfig}b and  Fig.\ref{mainfig}c suggests 
that luminous cool galaxies, where older stars have greater spatial 
distribution, show a tendency to have cool dust distributed over larger volume. 
This may be interpreted, as an indirect but interesting support, that FIR light 
from cool galaxies are reprocessed emission of photons coming predominantly 
from the old population of stars. This is in accord with the expectation that 
dust heated by the non-ionizing photons from a cooler population of stars might 
follow the smoothed distribution of older stars in the galaxy. 
In spite of a large scatter, we can see a moderate trend where warm sources 
have relatively smaller regions of NIR emission than their cool counterparts. 
Most of the ULIRGs (4/7) have larger NIR and $B$-band diameters than found among 
warm ($C > -0.2$) LIRGs (Fig.2c and 2d). It is interesting that ULIRGs, which 
all have very warm dust ($C > -0.1$), have NIR and B diameters comparable to 
the subset of LIRGs with relatively cool dust temperatures ($C < -0.22$). The 
data indicate that the ULIRGs have relatively large disks (e.g., they involve 
mergers of massive galaxies) with compact cores that dominate the FIR emission.
It is also clear from Fig.2c that warm LIRGs (large open circles) have 
systematically smaller NIR diameters than cool LIRGs (large solid circles).
We conclude that some galaxies with relatively cool dust temperatures
are LIRGs (rather than lower luminosity SFGs)  because, despite having a lower 
average SFR per unit area, they have on average larger total surface areas than 
warm LIRGs.

The Optical size, representing the spatial distribution of stellar emission, 
shows no trend with the color when the whole sample is considered 
(Fig.\ref{mainfig}d). However it improves for SFGs+LIRGs, after removing ULIRGs 
from the sample (see Table 2 for correlation statistics). The scatter 
is large as compared to Fig.\ref{mainfig}b and Fig.\ref{mainfig}c. All ULIRGs 
and a substantial number of warm SFGs and LIRGs are almost similar in optical 
size compared to their cool counterparts. All of these warm sources show signs 
of interaction such as tidal tail, disturbed outer region or presence of 
neighbors.  

With the exception of color and SB$_{1.49 GHz}$ all of the physical parameters 
depend on the distance. Figure \ref{mainfig} demonstrates that color is tightly 
correlated with distance independent measures. On the other hand correlation 
between color and galaxy size is weak. There are at least three factors that can 
contribute to the broader distribution in color-size correlations (see Table 2 
for correlation statistics). 
First, there may be large error in the distance measurement in spite of 
correction for non-Hubble flows since correlation between color vs. NIR or 
B-band surface brightness is relatively stronger compared to the respective 
sizes. 
Second, optical diameter $D_B$ of interacting/merging galaxies will overestimate 
the actual size because of the inherent difficulty associated with size estimates 
of these sources. Removing the ULIRGs from the sample makes the color-size trend 
stronger.
Third, varying scatters in color vs. size plots reflect methodological 
differences in estimating angular size in different wavelengths. The scatter 
would have been reduced if galaxy sizes were estimated in all wavelengths in 
a systematic manner, e.g., at the same brightness level.
% One may ask whether or not the tight correlation between 
%$C$ vs. SB$_{1.49 GHz}$ is biased: cool objects are mostly face on disks and 
%hence have larger area than the warmer sources which are highly edge on. The 
%sample galaxies of any category do not show any preferred location in the 
%color-inclination plot (not shown here). It suggests that galaxy inclination 
%has less significance in our analysis and we have an unbiased presentation of 
%physical characteristics of cool and warm IR sources.  

\begin{figure}
\epsscale{0.85}
\plotone{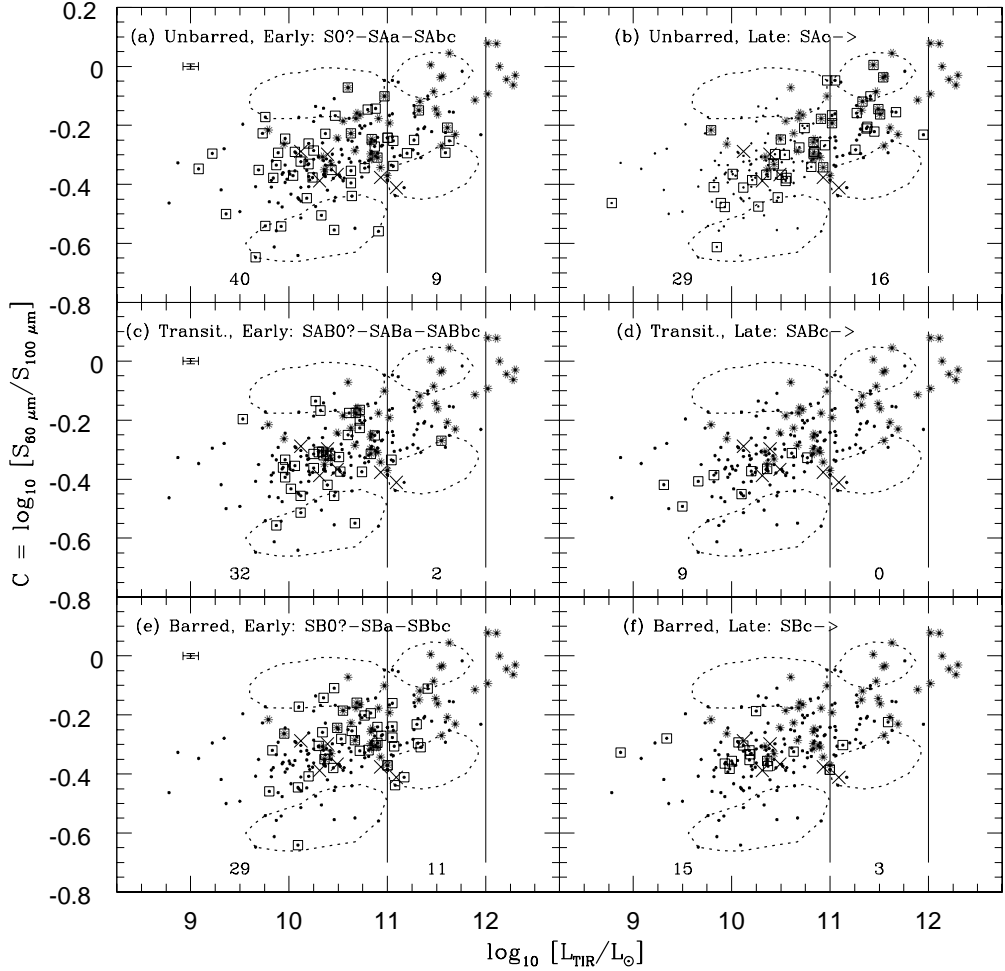}
\caption{FIR color as a function of galaxy (optical) morphology. 
Dotted closed curves show the regions of cool and warm sources. Large crosses 
and asterisks represent, respectively, local sub-mm and interacting/disturbed 
galaxies. Representative error bar is shown at the top left and the number of 
galaxies with different morphologies are shown at the bottom of each panel. 
Only those galaxies, 154 SGFs and 41 LIRGs, that have morphological 
classification available in RC3 catalog are shown. 
Note from the figure that both transitional and barred spirals appear to be 
low luminosity star forming galaxies whereas high luminosity warm galaxies 
are predominantly unbarred (i.e. normal spirals). Cool and warm sources, in 
a given luminosity class, contain spirals with various morphologies. 
\label{morpfig}}
\end{figure}

Figure \ref{morpfig} shows color as a function of galaxy (optical) morphology. 
We use RC3 classification and divide galaxies into three broad categories: 
unbarred (A), transitional (AB), and barred (B). 
In each morphology bin, we take S0?-S0a-Sa-Sb-Sbc sources as early type, 
and S?-Sc-Scd-Sd-I-Pec sources as late type. Early and late type division is 
made to be consistent with earlier studies that have found fundamental 
differences in the properties of early and late type galaxies, especially, 
barred galaxies (Combes \& Elmagreen 1993; Ho, Filippenko \& Sargent 1997; 
Sakamato et al. 1999; Sheth et al. 2005). 

We did not find any classification for 11 LIRGs and it is also uncertain 
for the seven ULIRGs [{\it See Table 1 in the electronic edition of the Journal for these 
galaxies}]. As a result, Fig.\ref{morpfig} shows 154 SFGs and 
41 LIRGs divided into SA, SAB, and SB type sources. 
This figure illustrates that high luminosity LIRGs are mostly unbarred 
spirals (Fig.\ref{morpfig}a, b). The transitional and barred spirals, in 
general, fall into the class of low luminosity SFGs (Fig. \ref{morpfig}c,d 
and e,f). 
The essence of this figure are the followings: for SFGs, late type 
spirals stay close to the mean color (Fig.\ref{morpfig}d and f). 
Early type spirals, on the other hand, are responsible for the scatter in 
color-luminosity diagram (Fig.\ref{morpfig}a, c, and e). 
Interestingly, cool sources consist of early type unbarred and transitional 
galaxies whereas warm sources consist of early types of all three categories. 
In other words, low luminosity barred galaxies, in general, show a tendency 
to have higher color temperature.  
For LIRGs, late type unbarred galaxies are predominately warm 
(Fig.\ref{morpfig}b). This is not unexpected since late type galaxies are 
highly irregular systems with higher star formation rate which results in 
warmer color. Early type LIRGs are systematically cooler (Fig.\ref{morpfig}a, 
c, and e). Both transitional and barred spirals are extremely rare in this 
luminosity class. 

In a recent study, Sheth et al. (2005) showed that late types barred spirals 
are less centrally concentrated than early types and a significant subset of 
early type barred spirals have little or no gas within the bar region. 
This observation has been explained as a result of higher mass accretion rates 
in the past in early type barred spirals where the large amount of gas driven 
inward by the bar have already been converted into stars. 
They suggested that these galaxies are in the post-starburst phase. The IR 
luminosity of Sheth et al. sample of 44 galaxies span the range,
$10^{9} < L_{TIR}/L_{\odot} < 10^{11}$, which we call star forming galaxies 
(SFGs) in our sample. These galaxies fall in the class of SFGs in our sample. 
The observation that early type barred spirals are in the post-starburst phase 
gives a plausible explanation for these galaxies to be low luminosity star 
forming systems.

Note that the RC3 classifications used here are based on optical images
that are strongly affected by dust obscuration in our FIR-selected
galaxy sample. More recent studies at NIR wavelengths indicate the presence
of stellar bars in a large fraction of disk galaxies that  appear unbarred
at optical wavelengths (Eskridge et al. 2000). In addition, when one extends 
the  bar detection sensitivity to a low relative amplitude of $\sim 3\%$ in 
the $K$ band,  nearly $90\%$  of a sample of optically unbarred (SA) spirals 
contain stellar bars (Grosbol, Patsis, \& Pompei 2004).
Therefore, the fractions of transition (SAB) and barred (SB) galaxies
is likely to be much higher than is indicated by the RC3 classifications,
and the optical bar fractions may be telling us more about the relative
extinction by dust at optical wavelengths than the intrinsic stellar
distributions. An improved understanding of this issue requires detailed 
analysis of near-IR images for a large sample of LIRGs and ULIRGs.
%%%%%%%%%%%%%%%%%%%%%%%%%%%%%%%%%%%%%%%%%%%%%%%%%%%%%%%%%%%%%%%%%%%%%%%%%%%%%%%%%%%%%
%%%%%%%%%%%%%%%%%%%%%%%%%%%%%%%%%%%%%%%%%%%%%%%%%%%%%%%%%%%%%%%%%%%%%%%%%%%%%%%%%%%%%
%%%%%%%%%%%%%%%%%%%%%%%%%%%%%%%%%%%%%%%%%%%%%%%%%%%%%%%%%%%%%%%%%%%%%%%%%%%%%%%%%%%%%
%%%%%%%%%%%%%%%%%%%%%%%%%%%%%%%%%%%%%%%%%%%%%%%%%%%%%%%%%%%%%%%%%%%%%%%%%%%%%%%%%%%%%
\section{Discussions}
It is necessary to address whether the large scatter in the FIR 
color-luminosity diagrams is real (intrinsic to the physics of the objects), 
or an artifact of the observations. Figure \ref{colufig} shows the data 
from the IRAS sample of Fisher et al. ($S_{60\mu m} > 1.2$ Jy;  panel a) 
and data from the RBGS ($S_{60\mu m} > 1.2$ Jy;  panel b). The main objective 
of the RBGS was to produce more accurate fluxes for the many large, nearby 
galaxies resolved by IRAS, by recovering extended emission that was not 
represented in the (underestimated) flux densities reported for such objects 
in the IRAS Point Source Catalog (PSC) and Faint Source Catalog (FSC). The 
RBGS also added some objects missed in the previous BGS compilations.
This problem is minimal for the fainter objects in the 1.2 Jy sample, which 
on average are more distant and therefore smaller than the RBGS objects, with 
little or no flux missed in the published PSC/FSC flux densities. The main 
conclusion we can draw from comparing Fig. 1a with Fig. 1b is that the scatter 
in the 5.24 Jy RBGS  is consistent with the scatter in the Fisher et al. 1.2 
Jy sample.
The presence of more true statistical outliers ($\geq 3 \sigma$ in each
luminosity bin) in the 1.2 Jy sample compared to the 5.24 Jy sample is 
expected from basic sampling statistics, as is the extension to higher 
$L_{TIR}$ values due to sampling of a much larger volume of space in the 1.2 
Jy sample. As stated in previous sections, the uncertainty in the $L_{TIR}$ 
values are dominated by the distance determinations ($\sim 15\%$), and the 
uncertainty in $C$ is relatively small (See the representative error bar in 
Fig. 1a). Therefore, we conclude that the dispersion in the $C-L_{TIR}$ 
digram is intrinsic and related to the diversity of physical conditions of 
the ISM in the galaxies.

%%%% describe nature of dust distribution and energy source  
To understand the physical origin of the color-luminosity diagram it is 
necessary  to have detailed knowledge of these cool and warm galaxies. 
Questions can be raised such as, in a given luminosity range, what are the 
possible (external or internal) mechanisms that would cause some galaxies 
to have cool color temperature compared to the majority of the galaxies ? 
what are the basic connections between the photometric and structural 
properties of galaxies in this diagram ?
We proceed in light of above queries and find that dust in the cool objects 
is less centrally concentrated than in the more typical LIRGs and ULIRGs 
that are dominated by more centrally concentrated warm dust. 
We also find that the optical disks of IR cool galaxies show a tendency to 
be more extended than those of the warmer ones. The color-size trend is 
relatively stronger at longer wavelengths. The extended emission in cool 
sources may well indicate heating by old stars, but it could also indicate 
simply wide-spread small star forming regions scattered in a large dusty 
disk. Distinguishing between these aspects is beyond the scope of this 
study.

The color-luminosity trend shown in Figure \ref{colufig} is observed in 
the local universe ($z < 0.082$). It had been demonstrated that the trend 
does not change out to $z \sim 1$ (Chapman et al. 2003) meaning that the 
dispersion of color in each luminosity bin does not vary significantly 
with redshift. However, given the presence of higher luminosity objects 
such as hyper-luminous galaxies with $L_{TIR}/L_{\odot} > 10^{13}$ in 
high-redshift samples that cover much larger volumes of space and earlier 
look-back times than local samples, from the present data we can only make
such inferences for objects with $L_{TIR}/L_{\odot} \leq 10^{12.5}$.
In this respect, this correlation, therefore, reveals something quite 
profound: in absence of any structural information of high $z$ luminous 
galaxies we can get a rough idea about the extent of the disks if we 
simply know the FIR colors and luminosities of these galaxies. In spite 
of a large scatter, the trend suggests that cool galaxies are generally 
highly luminous because of extended disks. The warm galaxies may or may 
not have extended disks depending on galaxy environment and/or internal 
processes.

Current sub-millimeter/millimeter (sub-mm/mm) surveys have discovered a 
new population of $z > 1$ sub-mm galaxies, contributing up to 50$\%$ of 
the extragalactic background light (Smail, Ivison, \& Blain 1997; Barger 
et al. 1998; Hughes et al. 1998; Borys et al. 2003). There are two possible 
explanations of these galaxies: they are either proto-spheroids with high 
star formation rate $\geq 10^3$ M$_{\odot}$ yr$^{-1}$ (Dunlop 2001) or 
disks dominated in the sub-millimeter by infrared cirrus heated by their 
interstellar radiation field, rather than intense star formation 
(Rowan-Robinson 2001; Efsthathiou \& Rowan-Robinson 2003, hereafter ERR; 
King \& Rowan-Robinson 2003). ERR have modeled cirrus emission for a 
selection of local ($z <0.02$) and high-$z$ ($>1$) sub-mm galaxies and  
found excellent agreement in all cases. Their work suggests that sub-mm 
count and background can be understood in terms of cirrus-like emission 
(with effective dust temperature $<30$ K) 
rather than invoking dusty ultraluminous starburst galaxies, e.g. Arp 220 
in our local universe, to be the representative of distant sub-mm galaxies.
The significance of ERR work is that if the cool component dominates the 
observed sub-mm fluxes of these high-z sub-mm/mm wave survey sources the 
emission should be extended rather then compact (i.e. centrally condensed).
We attempt to explore this aspect of sub-mm galaxies in the context of our 
study by analyzing the $1.49 \ GHz$ emission of the local sample of ERR 
since only these galaxies are available in the RBGS. 

The local sample of ERR includes UGC 0903, NGC 0958, NGC 1667,  NGC 2990, 
UGC 5376, NGC 5962, NGC 6181. We remove NGC 2990 from the ERR sample 
because it has $S_{60 \mu m} < 5.24$ Jy placing it below the flux limit 
of the RBGS. Except NGC 1667 and  UGC 5376, the other four galaxies are 
present in our sample. The IR and radio information of these two galaxies 
are obtained respectively from the RGBS and the NRAO VLA Sky Survey catalog 
(NVSS; Condon et al. 1998). The FIR color and total IR luminosities of 
these six galaxies (shown by large crosses) are similar to normal SFGs 
and cool LIRGs in our sample (Fig.1b). 
We find that these galaxies indeed have large radio (FIR) disks.
The $1.49 \ GHz$ radio source diameters of these galaxies range $\sim 4$ 
to 20 kpc, which is comparable to cool LIRGs and even normal SFGs (those 
not defined in this study as unusually cool or warm). This is in stark 
contrast to the warm SFGs and LIRGs that have radio sizes less than 
$\sim 3$ kpc (See Fig.2b).

The FIR cool high-$z$ sub-mm galaxies of ERR sample carry a few 
$\times 10^{13} L_{\odot}$, several hundred times more luminosity than 
these local examples. Assuming that these galaxies are scaled-up versions 
of local cool examples, where disk size increases at constant surface 
brightness to account for the increased luminosity, we find that their 
radio (and FIR) disks must range up to 100 kpc or more in size.
While rare, local galaxies with similar sizes do exist, as in the case of 
low surface-brightness, HI-rich galaxies (Sprayberry et al. 1995; Bothun, 
Impey, \& McGaugh 1997; Matthews, van Driel, \& Monnier-Ragaign 2001). At 
earlier epochs the cool sub-mm galaxies may be single large disks able to 
regulate their star formation at the modest level required for these 
luminosities and FIR colors. But what do we know from observations about 
the extent of FIR/Radio emission in these distant  sub-mm sources, 
especially the FIR cool ones ?

For the assumed cosmology in this study, the linear disk size of 100 kpc 
corresponds to an angular size of $> 7^{''}$ for $z > 1$ sub-mm galaxies. 
On the other hand, the reported sizes of sub-mm galaxies in the redshift 
range $1 \lesssim z \lesssim 4$ are significantly smaller, 
$\lesssim 1\acute{\acute{.}}5$ (Ivison et al. 2002; Iono et al. 2006;  
Tacconi et al. 2006). 
This suggests that the objects measured are best understood as scaled-up 
versions of local FIR warm ULIRGs rather than cool extended disks (Tacconi 
et al. 2006), since nothing is known about their FIR colors. While cold 
ULIRGs had been detected in the redshift range $0.4 < z < 1$, very little 
is known about their morphology or geometry (Chapman et al. 2002). However, 
the recently reported optical properties of BzK-15504 (Genzel et al. 2006) 
make it a very likely FIR cool luminous galaxy. At $z=2.38$, its disk 
extends out to $\sim 10$ kpc, and has elevated distributed star formation 
throughout in addition to an accreting central black hole. The estimated 
total star formation rate for this galaxy is $140^{+100}_{-80}$ M$_{\odot}$ 
yr$^{-1}$ (Genzel et al. 2006) suggesting that it is an ultra-luminous 
infrared galaxy with $L_{TIR} \gtrsim 10^{12}$ L$_{\odot}$.

The relative frequency of occurrence of FIR cool objects as a function of 
redshift is still unknown, and the detection efficiency will decrease as 
the source extent increases.
%The Tacconi et al. results only place a limit of X\% on the frequency
%of cool objects, since they have measured 8 objects, all of which
%were compact.  
Tacconi et al. have measured 8 objects, all of which were compact. Their 
results only place a limit on the effective dust temperature ($39 \pm 3$ 
K for $\beta = 1.5$) of sub-mm galaxies rather than the frequency of cool 
objects. As precise redshifts and more accurate 
rest frame FIR fluxes of high-$z$ galaxies become available, placing these 
sub-mm/mm galaxies in the context of the FIR color-luminosity diagram will 
help shed light on their structure. Combining that with kinematic information  
may constrain further the nature and evolving path of these galaxies.  

%%%% describe hubble tines
Warm LIRGs and ULIRGs are usually associated with intense bursts of star
formation. These galaxies are known to have higher star formation rates 
(SFR) and star formation efficiency (SFE) than lower luminosity galaxies.
These parameters are defined in such a 
manner that the SFR is directly related to the $L_{TIR}$ (Kennicutt 1998) 
whereas the SFE is the $L_{TIR}$ normalized by the molecular hydrogen mass 
(Young 1999). It has been noted that along the Hubble sequence, the SFR 
increases toward late-type, which could be due to dynamical instability 
increasing with decreasing bulge-to-disk ratio (Combes 2001). 
This trend is generally observed among the optically unbarred spirals in 
the present study as well, whereas galaxies with warmer FIR colors and 
hence higher SFRs are less likely to be transitional (SAB) or barred (SB) 
galaxies (See Fig.2)

%%%% describe morphology 
%%%% describe the physical processes/parameters connecting morphology and 
%%%% color-luminosity diagram 
Question can also be raised such as, what types of morphological features do 
the cool and warm galaxies have and how could these features be linked with 
the FIR color-luminosity correlation ? We find that the LIRGs in our sample 
are mostly late type unbarred spirals. The transitional and barred spirals, 
in general, fall into the class of low luminosity SFGs. Among LIRGs, the late 
type unbarred spirals are predominately warm, and the early type unbarred and 
barred are systematically cooler. One probable reason for the transitional 
and barred spirals being low luminosity star forming galaxies is that they 
are in the post-starburst phase, where the large amount of gas driven inward 
by the bar have already been converted into stars, consistent with the findings 
of Sheth et al. (2005). 
%For low luminosity unbarred galaxies, however, dust may be expelled toward or out 
%from the galaxy center by various mechanisms such as convective flows, radiation 
%pressure, turbulent motions etc. (Norman \& Ikeuchi 1989; Ferrara et al. 1991). 
%Interstellar dust has been recognized to spread over large distances from their 
%sources, and thus can be considered as a tracer of dynamical processes responsible 
%for the circulation of material in galaxy disk (Alton et al. 1998; Trewhella et al. 
%2000).

%%%%
%%%%
%%%% Is there any selction bias ? what about completeness of the sample ?
%%%%

Finally we note that the low luminosity early type unbarred (SA) and 
transitional (SAB) spirals are responsible for the large scatter in the 
$C-L_{TIR}$ diagram. 
It may be that, in spite of our selection process, we have dust enshrouded 
AGN in the sample. These sources could span the entire luminosity range 
considered and could contribute to the scatter in this diagram. On the other 
hand this scatter may also arise from more variability in the SFR and star 
formation behavior, e.g., short episodic starburst or more possible 
configurations, e.g., large bulges and small star formation disks of early 
type sources.
%%%%%%%%%%%%%%%%%%%%%%%%%%%%%%%%%%%%%%%%%%%%%%%%%%%%%%%%%%%%%%%%%%%%%%%%%%%%%%%%%%%%%
%%%%%%%%%%%%%%%%%%%%%%%%%%%%%%%%%%%%%%%%%%%%%%%%%%%%%%%%%%%%%%%%%%%%%%%%%%%%%%%%%%%%%
%%%%%%%%%%%%%%%%%%%%%%%%%%%%%%%%%%%%%%%%%%%%%%%%%%%%%%%%%%%%%%%%%%%%%%%%%%%%%%%%%%%%%
%%%%%%%%%%%%%%%%%%%%%%%%%%%%%%%%%%%%%%%%%%%%%%%%%%%%%%%%%%%%%%%%%%%%%%%%%%%%%%%%%%%%%
%%%%%%%%%%%%%%%%%%%%%%%%%%%%%%%%%%%%%%%%%%%%%%%%%%%%%%%%%%%%%%%%%%%%%%%%%%%%%%%%%%%%%
%%%%%%%%%%%%%%%%%%%%%%%%%%%%%%%%%%%%%%%%%%%%%%%%%%%%%%%%%%%%%%%%%%%%%%%%%%%%%%%%%%%%%
\section{Conclusions}
We have analyzed a sample of 213 nearby IR galaxies to study the correlations 
of galaxy FIR color vs. several photometric and physical parameters. The set 
of measures include total IR luminosity, radio surface brightness as 
well as radio, NIR, and optical sizes. Our objective is to understand cool and 
warm IR sources using various correlations. 
We find that galaxy radio surface brightness is well correlated with FIR color 
whereas size is less well correlated with color. A weak color-size correlation 
signals a significant uncertainty associated with the distance measurements of 
sample galaxies. It may also reflect methodological differences in estimating 
angular size in different wavelengths. 
%%%%
We also find that late type galaxies (from Sc and beyond) of all morphology 
classes show less scattering than early types in the color-luminosity diagram.  
Our study shows that dust in the cool IR sources is probably less centrally 
concentrated than in the more typical luminous and ultra-luminous IR galaxies 
that are dominated by more centrally concentrated warm dust. We believe this 
result has significant implication in terms of FIR color-luminosity diagram: in 
absence of any structural information of high-$z$ luminous sources one can get 
rough estimate of the extent of the disks if one simply knows the FIR colors 
and luminosities of galaxies.
%%%%%%%%%%%%%%%%%%%%%%%%%%%%%%%%%%%%%%%%%%%%%%%%%%%%%%%%%%%%%%%%%%%%%%%%%%%%%%%%%%%%%
%%%%%%%%%%%%%%%%%%%%%%%%%%%%%%%%%%%%%%%%%%%%%%%%%%%%%%%%%%%%%%%%%%%%%%%%%%%%%%%%%%%%%
%%%%%%%%%%%%%%%%%%%%%%%%%%%%%%%%%%%%%%%%%%%%%%%%%%%%%%%%%%%%%%%%%%%%%%%%%%%%%%%%%%%%%
%%%%%%%%%%%%%%%%%%%%%%%%%%%%%%%%%%%%%%%%%%%%%%%%%%%%%%%%%%%%%%%%%%%%%%%%%%%%%%%%%%%%%
%%%%%%%%%%%%%%%%%%%%%%%%%%%%%%%%%%%%%%%%%%%%%%%%%%%%%%%%%%%%%%%%%%%%%%%%%%%%%%%%%%%%%
%%%%%%%%%%%%%%%%%%%%%%%%%%%%%%%%%%%%%%%%%%%%%%%%%%%%%%%%%%%%%%%%%%%%%%%%%%%%%%%%%%%%%
\acknowledgments
We thank the referee Jim Houck for insightful comments, and Thomas Jarrett 
for many useful communications.
NR thanks Ranga Ram Chary, Roc Cutri, Justin Howell, Guilain Lagache, Seppo 
Laine, Naveen Reddy, Kevin Xu, Min Su Yun, Zang Wang for discussions. NR 
gratefully acknowledges the support of a Research Associateship administered 
by the National Research council (NRC; upto December, 2005) and currently by 
the Oak Ridge Associated Universities (ORAU) during this research. 
This study has made use of the NASA/IPAC Extragalactic Database (NED) which is 
operated by the Jet Propulsion Laboratory, California Institute of Technology, 
USA under contract with the National Aeronautics and Space Administration, and 
the LEDA database in France. 
%(http://leda.univ-lyon1.fr).
%%%%%%%%%%%%%%%%%%%%%%%%%%%%%%%%%%%%%%%%%%%%%%%%%%%%%%%%%%%%%%%%%%%%%%%%%%%%%%%%%%%%%
%%%%%%%%%%%%%%%%%%%%%%%%%%%%%%%%%%%%%%%%%%%%%%%%%%%%%%%%%%%%%%%%%%%%%%%%%%%%%%%%%%%%%
%%%%%%%%%%%%%%%%%%%%%%%%%%%%%%%%%%%%%%%%%%%%%%%%%%%%%%%%%%%%%%%%%%%%%%%%%%%%%%%%%%%%%
%%%%%%%%%%%%%%%%%%%%%%%%%%%%%%%%%%%%%%%%%%%%%%%%%%%%%%%%%%%%%%%%%%%%%%%%%%%%%%%%%%%%%
%%%%%%%%%%%%%%% toy models

%%%%%%%%%%%%%%%%%%%%%%%%%%%%%%%%%%%%%%%%%%%%%%%%%%%%%%%%%%%%%%%%%%%%%%%%%%%%%%%%%%%%%
%%%%%%%%%%%%%%%%%%%%%%%%%%%%%%%%%%%%%%%%%%%%%%%%%%%%%%%%%%%%%%%%%%%%%%%%%%%%%%%%%%%%%
%%%%%%%%%%%%%%%%%%%%%%%%%%%%%%%%%%%%%%%%%%%%%%%%%%%%%%%%%%%%%%%%%%%%%%%%%%%%%%%%%%%%%
%%%%%%%%%%%%%%%%%%%%%%%%%%%%%%%%%%%%%%%%%%%%%%%%%%%%%%%%%%%%%%%%%%%%%%%%%%%%%%%%%%%%%
{}
%%%%%%%%%%%%%%%%%%%%%%%%%%%%%%%%%%%%%%%%%%%%%%%%%%%%%%%%%%%%%%%%%%%%%%%%%%%%%%%%%%%%%%
%%%%%%%%%%%%%%%%%%%%%%%%%%%%%%%%%%%%%%%%%%%%%%%%%%%%%%%%%%%%%%%%%%%%%%%%%%%%%%%%%%%%%%
%%%%%%%%%%%%%%%%%%%%%%%%%%%%%%%%%%%%%%%%%%%%%%%%%%%%%%%%%%%%%%%%%%%%%%%%%%%%%%%%%%%%%%
%%%%%%%%%%%%%%%%%%%%%%%%%%%%%%%%%%%%%%%%%%%%%%%%%%%%%%%%%%%%%%%%%%%%%%%%%%%%%%%%%%%%%%
%%%%%%%%%%%%%%%%%%%%%%%%%%%%%%%%%%%%%%%%%%%%%%%%%%%%%%%%%%%%%%%%%%%%%%%%%%%%%%%%%%%%%%
%%%%%%%%%%%%%%%%%%%%%%%%%%%%%%%%%%%%%%%%%%%%%%%%%%%%%%%%%%%%%%%%%%%%%%%%%%%%%%%%%%%%%%
%%%%%%%%%%%%%%%%%%%%%%%%%%%%%%%%%%%%%%%%%%%%%%%%%%%%%%%%%%%%%%%%%%%%%%%%%%%%%%%%%%%%%%
%%%%%%%%%%%%%%%%%%%%%%%%%%%%%%%%%%%%%%%%%%%%%%%%%%%%%%%%%%%%%%%%%%%%%%%%%%%%%%%%%%%%%%
%%%% Table 1
\begin{deluxetable}{ccccccccccccccccc}
\rotate
\tabletypesize{\tiny}
\tablewidth{0pc}
\setlength{\tabcolsep}{0.05in} 
\tablecolumns{17}
\tablecaption{Partial list of low luminosity star forming galaxies (SFGs)}
\tablehead
{
\colhead{Num.}   &\colhead{Name} &\colhead{$S_{12}$} &\colhead{$S_{25}$} 
&\colhead{$S_{60}$} &\colhead{$S_{100}$} &\colhead{$S_{R}$} 
&\colhead{$d$} 
&\colhead{Flag} 
&\colhead{$L_{TIR}$} &\colhead{$\theta^{B}_M$} &\colhead{$\theta^B_m$}  
&\colhead{$\theta^{NIR}_M$} &\colhead{$\theta^{R}_M$} 
&\colhead{$\theta^{R}_m$}   
&\colhead{$\theta^{R}_{beam}$} 
&\colhead{Morp.} \\ 
1   &\colhead{2}  &\colhead{3}  &\colhead{4}  &\colhead{5}  
&\colhead{6}      &\colhead{7}  &\colhead{8}  &\colhead{9}  
&\colhead{10}     &\colhead{11} &\colhead{12} &\colhead{13} 
&\colhead{14}     &\colhead{15} &\colhead{16} &\colhead{17}     
} 
%%%%
\startdata
 1   &NGC4565$^c$  &1.36 &1.36  &7.79  &34.62  &18.0  &9.99 &S  &9.66 &954.0 &114.0 &191.1 &360.0 &42.0 &60.0 &SA(s)b?  \\
 2   &NGC3953$^c$  &1.10 &1.19  &7.11  &31.12   &7.2 &17.58 &S &10.09 &414.0 &210.0 &132.3 &210.0 &84.0 &60.0 &SB(r)bc  \\
 3   &NGC5907$^c$  &1.29 &1.44  &9.14  &37.43  &16.2 &12.08 &S  &9.85 &768.0  &78.0 &235.3 &216.0 &36.0 &48.0 &SA(s)c:  \\
 4   &NGC3147$^c$  &1.95 &1.03  &8.17  &29.61  &49.8 &41.41 &- &10.91 &234.0 &210.0  &83.1  &60.0 &54.0 &60.0 &SA(rs)bc \\
 5   &NGC4579$^c$  &1.12 &0.78  &5.93  &21.39  &62.9 &15.29 &V  &9.87 &354.0 &282.0 &120.0  &54.0 &24.0 &54.0 &SAB(rs)b \\
 6   &NGC0772$^c$  &1.10 &0.92  &6.73  &24.15  &27.2 &28.71 &S &10.46 &432.0 &258.0 &105.2  &72.0 &66.0 &60.0 &SA(s)b   \\
 7   &NGC5371$^c$  &0.86 &0.97  &5.27  &18.66  &10.3 &41.06 &- &10.67 &264.0 &210.0 &102.4 &120.0 &96.0 &54.0 &SAB(rs)bc\\
 8   &NGC3675$^c$  &1.43 &1.67 &10.48  &36.56  &17.2 &12.69 &S  &9.92 &354.0 &186.0 &120.7  &84.0 &36.0 &48.0 &SA(s)b   \\
 9   &NGC4013$^c$  &0.54 &0.77  &7.01  &24.36  &13.9 &13.76 &S  &9.76 &314.9  &63.0 &111.4  &90.0 &12.0 &48.0 &SAb      \\
 10  &NGC4699$^c$  &0.76 &0.54  &6.11  &19.95   &8.5 &21.71 &- &10.12 &228.0 &156.0 &100.9  &60.0 &48.0 &60.0 &SAB(rs)b \\
 11  &NGC4501$^c$  &2.29 &2.98 &19.68  &62.97  &73.7 &15.29 &V &10.33 &414.0 &222.0 &155.4 &120.0 &66.0 &54.0 &SA(rs)b  \\
 15  &NGC0908$^c$  &1.74 &2.21 &17.54  &52.35  &35.8 &15.75 &S &10.27 &360.0 &156.0 &128.9  &90.0 &66.0 &48.0 &SA(s)c   \\
 20  &NGC5005$^c$  &1.65 &2.26 &22.18  &63.40  &49.5 &18.09 &- &10.46 &348.0 &168.0 &130.8  &96.0 &36.0 &48.0 &SAB(rs)bc\\
 24  &NGC3672$^c$  &1.01 &0.95  &9.23  &25.69  &23.3 &27.70 &- &10.47 &252.0 &114.0  &94.0  &84.0 &42.0 &48.0 &SA(s)c   \\
 25  &NGC4030$^c$  &1.35 &2.30 &18.49  &50.92  &66.5 &24.50 &- &10.64 &252.0 &180.0  &77.4  &72.0 &60.0 &60.0 &SA(s)bc  \\
 140 &NGC4102$^w$  &1.77 &6.83 &46.85  &70.29  &46.1 &16.89 &- &10.61 &162.0  &60.0  &68.6   &3.3  &2.2  &1.5 &SAB(s)b? \\
 141 &NGC7465$^w$  &0.26 &0.67  &5.47   &8.14  &11.7 &27.44 &- &10.10  &72.0  &48.0  &27.8   &6.0  &5.0 &15.0 &SB(s)0   \\
 142 &NGC4383$^w$  &0.29 &1.08  &8.36  &12.43   &4.8 &15.29 &V  &9.76 &114.0  &60.0  &35.0   &8.0  &5.0  &5.0 &SAa?     \\
 143 &NGC4536$^w$  &1.55 &4.04 &30.26  &44.51  &36.1 &14.92 &P &10.32 &456.0 &192.0 &113.5  &11.0  &6.0  &5.0 &SAB(rs)bc\\
 144 &NGC3471$^w$  &0.33 &1.26  &8.31  &12.21  &12.2 &34.07 &- &10.47 &104.4  &50.4  &35.2   &8.0  &5.0  &6.0 &SAa      \\
 145 &NGC5930$^w$  &0.35 &1.60  &9.36  &13.68  &10.2 &42.47 &- &10.72 &102.0  &54.0  &46.2   &2.8  &1.8  &1.5 &SAB(rs)b \\
 146 &NGC2798$^w$  &0.76 &3.21 &20.60  &29.69   &9.9 &27.84 &- &10.69 &156.0  &60.0  &38.7   &4.5  &1.6  &1.5 &SB(s)a   \\
 147 &NGC1482$^w$  &1.55 &4.68 &33.36  &46.73  &17.8 &25.09 &- &10.80 &150.0  &84.0  &42.9   &8.9  &4.2  &2.1 &SA0+     \\
 148 &NGC1204$^w$  &0.25 &1.10  &7.33  &10.18  &14.2 &58.51 &- &10.88  &66.0  &18.0  &33.3   &1.9  &1.0  &1.8 &S0/a:    \\
 149 &NGC1022$^w$  &0.71 &3.28 &19.71  &27.33  &26.4 &19.33 &- &10.35 &144.0 &120.0  &56.3   &4.0  &3.0  &6.0 &SB(s)a   \\
 150 &NGC3885$^w$  &0.57 &1.47 &11.89  &16.25  &23.4 &22.93 &- &10.27 &144.0  &60.0  &56.8   &8.0  &3.0  &7.0 &SAB(r:)0/a\\
 151 &NGC1266$^w$  &0.25 &1.20 &13.13  &16.89  &75.7 &28.86 &- &10.46  &90.0  &60.0  &32.8   &5.0  &2.0  &6.0 &SB(rs)0  \\  
 152 &NGC3597$^w$  &0.67 &2.18 &12.84  &16.21  &34.8 &48.31 &- &10.97 &114.0  &90.0  &25.0   &8.0  &5.0  &7.0 &S0+:     \\ 
 153 &NGC1222$^w$  &0.50 &2.28 &13.06  &15.41  &43.8 &32.26 &- &10.60  &66.0  &54.0  &22.9  &11.0  &7.0 &18.0 &S0-      \\  
 154 &NGC0839$^w$  &0.52 &2.27 &11.67  &13.03  &15.1 &51.10 &- &10.97  &84.0  &42.0  &29.2   &2.6  &1.3  &1.8 &SApec    \\
\hline
\enddata \\
\tablecomments{$c$ :cool; $w$ :warm; 
$\theta_M$ :major axis; $\theta_m$ :minor axis \\
Col.1 :Number; Col.2 :Galaxy Name;
Col.s 3-6 :$IRAS$ flux densities (Jy) from Revised Bright Galaxy Sample 
(RBGS; Sanders et al. 2003); 
Col.7 :$1.49 \ GHZ$ flux density (mJy) (Peak flux S$_P$ in Condon, 
Anderson \& Helou 1991; Wang \& Helou 1992) 
Col.8 : estimated galaxy distance $d$ in Mpc;
col.9 : flags P,S,V on distance estimate (see RBGS Table 1 for details); 
Col.10 :total IR luminosity $\log_{10}[L_{TIR}]$ in $L_{\odot}$ (RBGS);
Col.s 11-12:$B-$band 25 mag arcsec$^{-2}$ major and minor axes (arcsec);   
Col.13 :$K_s$ band 20 mag arcsec$^{-2}$ major axis (arcsec);
Col.14-15 :$1.49 \ GHZ$ major and minor axes for the resoluation at which 
the galaxy is resolved (arcsec); 
Col.16 :$1.49 \ GHZ$ beam resolutiton;
Col.17 : Galaxy optical morphology. 
}
\end{deluxetable} 
%%%%%%%%%%%%%%%%%%%%%%%%%%%%%%%%%%%%%%%%%%%%%%%%%%%%%%%%%%%%%%%%%%%%%%%%%%%%%%%%%%%%%%%
%%%%%%%%%%%%%%%%%%%%%%%%%%%%%%%%%%%%%%%%%%%%%%%%%%%%%%%%%%%%%%%%%%%%%%%%%%%%%%%%%%%%%%%
%%%%%%%%%%%%%%%%%%%%%%%%%%%%%%%%%%%%%%%%%%%%%%%%%%%%%%%%%%%%%%%%%%%%%%%%%%%%%%%%%%%%%%%
%%%%%%%%%%%%%%%%%%%%%%%%%%%%%%%%%%%%%%%%%%%%%%%%%%%%%%%%%%%%%%%%%%%%%%%%%%%%%%%%%%%%%%%
%%%% Table 2
\begin{deluxetable}{ccccccccc}
\tabletypesize{\small}
\tablewidth{0pc}
\setlength{\tabcolsep}{0.04in} 
\tablecolumns{9}
\tablecaption{Correlation statistics from Pearson's Correlation Test}
\tablehead{
\colhead{Galaxy Class}   
&\multicolumn{2}{c}{$C - SB_R$} 
&\multicolumn{2}{c}{$C - D_R$} 
&\multicolumn{2}{c}{$C - D_{NIR}$} 
&\multicolumn{2}{c}{$C - D_B$} \\ \hline
\colhead{}   
&\colhead{r}  &\colhead{$P$}  
&\colhead{r}  &\colhead{$P$}  
&\colhead{r}  &\colhead{$P$}  
&\colhead{r}  &\colhead{$P$} 
}
%%%%
\startdata
SFGs+LIRGs+ULIRGs 
        &0.77  &1.0E-08  &-0.65  &2.14E-07  &-0.14  &4.74E-02  &-0.11  &1.15E-01\\
SFGs+LIRGs
        &0.77  &1.0E-08  &-0.67  &1.83E-08  &-0.22  &1.91E-03  &-0.19  &5.33E-03\\
\hline
\enddata \\
\tablecomments{Correlation statistics for FIR color $C$ vs. all other 
parameters in Fig. \ref{mainfig}. In this table, $r$ is the correlation 
coefficient, and $P$ is the probability that correlation could arise from 
an uncorrelated sample. The confidance is $1-P$. 
Galaxy $1.49 \ GHz$ surface brightness and size are represented by $SB_R$ 
and $D_R$. For both $K_s$ and $B$ bands, test is performed after removing 
galaxies with unknown sizes.}
\end{deluxetable}

\end{document}